\documentclass[runningheads]{cl2emult}
\newcommand{\Ox}{Noe}
\usepackage{ifthen}
\usepackage{psfig}
\usepackage{epsfig}
\usepackage{epsf}
\usepackage{makeidx}  
\usepackage{graphicx} 
\usepackage{subeqnar} 
\usepackage{multicol} 
\usepackage{citesort}
\usepackage{amssymb}
\usepackage{latexsym}
\usepackage{cropmark} 
\usepackage{lnp}      
\usepackage{psfrag}  
\makeindex            

\ifthenelse{\equal{\Ox}{Ja}}{
\newcommand{\cal}{\mathcal}}

\begin{document}
\setcounter{tocdepth}{0}
\ifthenelse{\equal{\Ox}{Ja}}{

\frontmatter
\mainmatter}
\setcounter{page}{107}
\title*{\vspace*{-1.3cm}{\small In: J.A. Freund and T. P\"oschel (eds.), {\em Stochastic Processes in Physics,}}\vspace*{-0.25cm}\\ {\small {\em Chemistry, and Biology}, Lecture Notes in Physics Vol. 557, Springer}\vspace*{-0.25cm}\\ {\small (2000), p. 107}\vspace*{0.3cm}\\ Diffusion in Granular Gases\protect\newline of Viscoelastic Particles\vspace*{-0.25cm}}
\toctitle{Diffusion in granular gases of viscoelastic particles}
\titlerunning{Diffusion in granular gases of viscoelastic particles}
\author{Nikolai V. Brilliantov\inst{1}  \inst{2}
\and Thorsten P\"oschel\inst{3}}
\authorrunning{N.\,V. Brilliantov and T. P\"oschel}
\tocauthor{N.\,V. Brilliantov, T. P\"oschel}
\institute{Moscow State University, Department of Physics, 119899 Moscow, Russia
\and MPIKG, Am M\"uhlenberg, 14424 Potsdam, Germany
\and Universit\"at Stuttgart, Institute for Computer Applications, Pfaffenwaldring 27, D-70569 Stuttgart, Germany\vspace*{-0.25cm}}
\maketitle 

\begin{abstract} 
In most of the literature on granular gases it is assumed that the restitution coefficient $\epsilon$, which quantifies the loss of kinetic energy upon a collision is independent on the impact velocity. Experiments as well as theoretical investigations show, however, that for real materials the restitution coefficient depends significantly on the impact velocity. We consider the diffusion process in a homogeneous granular gas, i.e. in a system of dissipatively colliding  particles. We show that the mean square displacement of the particles changes drastically if we take the impact velocity dependence of $\epsilon$ into account. Under the oversimplifying assumption of a constant coefficient one finds that the particles spread in space logarithmically slow with time, whereas realistic particles spread due to a power law.
\end{abstract}

\section{Introduction}
Granular gases, i.e. gases composed of particles of mesoscopic or macroscopic size which suffer inelastic collisions, may be described under certain assumptions by the same quantities as common molecular gases. One can define for these systems temperature, density, velocity field, etc., and describe them  within the framework of hydrodynamics. Due to the inelasticity of the particles, however, there emerge new additional time and length scales. 

From a phenomenological point of view, granular gases behave very differently from molecular gases. One observes cluster formation,  e.g.,~\cite{GZ93,McNamara,Cluster,clus2} and vertexes, e.g.,~\cite{NEBO97} during the evolution of an initially uniform granular gas. Although plausible explanations of these effects exist,~\cite{GZ93,McNamara,Cluster,NEBO97} they still lack a comprehensive theory. 

Starting from a homogeneous state, all structures in granular gases develop after some time-lag. Before noticeable inhomogeneities appear the gas evolves in the homogeneously cooling state (HCS), when it gradually loses its energy according to inelastic collisions. During this regime of the evolution the gas is completely described by the current temperature (which decreases with time) and by the velocity distribution function. The HCS which precedes all further states of evolution, as e.g. clustering, is the most simple state of a granular gas. Understanding the physics of the HCS might, therefore, shed some light on more complex phenomena such as cluster and vortex formation.

In the present study we focus on the diffusion processes in granular gases being in the HCS. 
We consider the case of a constant restitution coefficient, which frequently is assumed in literature but turns out to be an oversimplified  model of a real collision, and impact-velocity dependent restitution coefficient as it is valid for viscoelastic particle interaction.

\section{Collision of Particles} 
The microscopic dynamics of granular particles is governed by the (normal) restitution coefficient $\epsilon$ which relates the normal components of the particle velocities before and after a collision, $\vec{v}_{\mbox{\footnotesize\em  ij}}\equiv\vec{v}_{i}-\vec{v}_{j}$ and $\vec{v}_{\mbox{\footnotesize\em ij}}^{\,\prime}\equiv\vec{v}_i^\prime-\vec{v}_j^\prime$ by $\left|\vec{v}_{\mbox{\footnotesize\em ij}}^{\,\prime} \vec{e}\right| = \epsilon \left|\vec{v}_{\mbox{\footnotesize\em ij}} \vec{e}\right|$. The unit vector $\vec{e}=\vec{r}_{\mbox{\footnotesize\em ij}}/|\vec{r}_{\mbox{\footnotesize\em ij}}|$ gives the direction of the inter-center vector $\vec{r}_{\mbox{\footnotesize\em ij}}=\vec{r}_i-\vec{r}_j$ at the instant of the collision. From the conservation of momentum one finds the change of velocity for the colliding particles:
\begin{equation}
\vec{v}_{i}^{\,\prime} = \vec{v}_{i} -\frac12 \,(1+\epsilon)\left(\vec{v}_{\mbox{\footnotesize\em ij}} \cdot  \vec{e}\,\right) \vec{e}\,,\,\,\,\,\,\,\,\,\,\,\,\,\,\,\,
\vec{v}_{j}^{\,\prime} = \vec{v}_{j} +\frac12 \,(1+\epsilon)\left(\vec{v}_{\mbox{\footnotesize\em ij}} \cdot  \vec{e}\,\right) \vec{e}\, .
\label{eq:restdef}
\end{equation}
For elastic collisions one has $\epsilon =1$ and for inelastic collisions $\epsilon $ decreases with increasing degree of inelasticity. 

In literature it is frequently assumed that the restitution coefficient is a material constant, $\epsilon=\mbox{const.}$ Experiments, e.g.~\cite{CollExp}, as well as theoretical investigations~\cite{BSHP} show, however, that this assumption is not consistent with the nature of the inelastic collisions, it does not agree even with a dimension analysis~\cite{Rozaetal}. The impact velocity dependence of the restitution coefficient $\epsilon\left(\left|\vec{e}\,
\vec{v}_{\mbox{\footnotesize\em ij}}\right|\right)$ has been obtained by generalizing Hertz's contact problem to viscoelastic spheres~\cite{BSHP}. From the generalized Hertz equation one obtains the velocity-dependent restitution coefficient~\cite{TomThor}
\begin{equation}
\epsilon=1- C_1\left(\frac{3A}{2} \right) \alpha^{2/5}\left|\vec{e}\,\vec{v}_{\mbox{\footnotesize\em ij}}\right|^{1/5} 
+C_2 \left(\frac{3A}{2} \right)^2  \alpha^{4/5}\left|\vec{e}\,\vec{v}_{\mbox{\footnotesize\em ij}}\right|^{2/5}  \mp  \cdots
\label{epsilon}
\end{equation}
with 
\begin{equation}
\alpha= 
\frac{ 2~ Y\sqrt{R^{\,\mbox{\footnotesize\em eff}}}}{
    3~ m^{\mbox{\footnotesize\em eff}}\left( 1-\nu ^2\right) }\,,
\end{equation}
where $Y$ is the Young modulus, $\nu$ is the Poisson ratio, and $A$
depends on dissipative parameters of the particle material 
(for details see~\cite{BSHP}). The effective mass and radius are defined as
\begin{equation}
R^{\,\mbox{\footnotesize\em eff}}=R_1R_2/(R_1+R_2)~~~~~~~~~~
m^{\mbox{\footnotesize\em eff}}=m_1m_2/(m_1+m_2)
\end{equation}
with $R_{1/2}$ and $m_{1/2}$ being the radii and masses of the
colliding particles. The constants 
are given by \cite{TomThor,Rozaetal}
\begin{equation}
C_1=\frac{ \Gamma(3/5)\sqrt{\pi}}{2^{1/5}5^{2/5} \Gamma(21/10)}\approx 1.1534 ~~~~~~~~~~
C_2=\frac35C_1^2\approx 0.7982 \,.   
\label{C2}
\end{equation}
Equation (\ref{epsilon}) refers to the case of pure viscoelastic
interaction, i.e. when the relative velocity $|\vec{v}_{ij}\vec{e}|$
is not too large (to avoid plastic deformation of the particles) and is 
not too small (to allow to neglect surface effects such as roughness,
adhesion and van der Waals interactions). 

\section{Granular Gas Dynamics}
\label{BP:GGD}
  
The diffusion equation
\begin{equation}
\vec{J}_A (\vec{r})= - D_A \vec{\nabla} n_A (\vec{r})
\end{equation}
describes the current $\vec{J}_A(\vec{r})$ of particles of sort $A$ at position $\vec{r}$ against the density gradient  $\vec{\nabla} n_A (\vec{ r})$ with $D_A$ being the diffusion coefficient. Using the continuity equation 
\begin{equation}
\frac{\partial n_A (\vec{r}) }{\partial t} =- \vec{\nabla} \vec{J}_A (\vec{r})
\end{equation}
we obtain the ``canonical'' diffusion equation:
\begin{equation}
\label{difcanon}
\frac{\partial n_A (\vec{r}) }{\partial t} = D_A \vec{\nabla}^2 n_A (\vec{r})\,.
\end{equation} 
Usually the term diffusion refers to the motion of a species of particles $A$ in a ``solution'' of other particles $B$. It is also possible, however, that $A$ and $B$ are of the same type, distinguishable only by a feature which does not affect the mechanical properties, i.e. mechanically $A$ and $B$ are indistinguishable; one may think of $A$ and $B$ having different colors. The process of diffusion of tagged particles among mechanically identical ones is called self-diffusion. 

The diffusion (and self-diffusion) coefficient is closely related with the mean square displacement of tagged particles with time. Assume the tagged particles are located at time $t=0$ in the origin $\vec{r}=0$. Then one can write for the square average displacement of particles at time $t$:
\begin{equation}
\label{sqavdef}
\left\langle \left[ \vec{r} (t) \right]^2  \right\rangle= \left.\int d \vec{r} \left[ \vec{r(t)}\right]^2 n_A (\vec{r},t) \right/\int d \vec{r} n_A (\vec{r}) 
\end{equation} 
where $ n_A (\vec{r},t) /\int d \vec{r} n_A (\vec{r},t)$ is the fraction of particles $A$ located at $\vec{r}$ at time $t$. (Note that $\int d \vec{r} n_A (\vec{r},t)=N={\rm const}$ is the total number of tagged particles.) Now we multiply both sides of Eq.~(\ref{difcanon}) with $\vec{r}^2$ and integrate over $d\vec{r}$. Using two times integration by parts and definition (\ref{sqavdef}), one obtains for 3D-systems
\begin{equation}
\label{sqavdef1}
\frac{d }{d t} \left\langle \left[ \vec{r} (t) \right]^2  \right\rangle =6D_A\, ;  \qquad \qquad \left\langle \left[ \vec{r} (t) \right]^2  \right\rangle =6D_At\,.
\end{equation} 
Using the kinematic relation ${\bf r}(t)= \int_0^t {\bf v}(t^{\prime}) dt^{\prime}$, one obtains
\begin{equation}
\left\langle \left[ \vec{r}(t) \right]^2  \right\rangle  =\left< \int_0^{t}\vec{v}(t^{\prime})dt^{\prime}\int_0^t  \vec{v}(t^{\prime \prime}) dt^{\prime \prime}  \right> 
\label{delR}
\end{equation}
and encounters with the velocity autocorrelation function (VAF) $\left< {\bf v}(t^{\prime}) {\bf v}(t^{\prime \prime}) \right> $.  
For gases in equilibrium the VAF depends only on the time difference, $|t^{\prime}-t^{\prime \prime}|$ and decays with a characteristic time $\tau_v$. Using these properties of the VAF, one can perform time-integration in (\ref{delR}) for $t\gg \tau_v$, i.e. for $t\rightarrow\infty$. Taking into account (\ref{sqavdef1}) one obtains the basic relation
\begin{equation}
D_A=\frac13 \, \int_0^{\infty}\,\left\langle \vec{v}(0) \vec{v} (t)\,\right\rangle dt \,.
\label{Dvel}
\end{equation}

Although not stated explicitly, all the above discussion refers to the case of equilibrium gases. Granular gases are {\em a priori} non-equilibrium systems, nevertheless, the concept of the diffusion coefficient may be generalized for such systems. Obviously, this refers only to ``liquid'' or gaseous phases of the material~\cite{EsipovPoeschel:97} where the particles are highly mobile. As discussed in the introduction throughout this article we assume that the granular material is homogeneous and isotropic (HCS). 

Another quantity which we will need below is the ``granular temperature'' $T(t)$, which decreases with time due to the loss of kinetic energy according to inelastic collisions. If the particles do not lose too much energy in a collision, the temperature changes on a time-scale $\tau_0$, which is much larger than the mean collision time $\tau_c$, i.e. $\tau_0 \gg \tau_c$~\cite{BrilliantovPoeschelPRE200:dif}. This condition allows for the definition of temperature, but imposes some restrictions for the intervals of time and values of material parameters. For a full discussion see~\cite{BrilliantovPoeschelPRE200:dif}. 

Whereas the diffusion coefficient $D$ for equilibrium systems is just a constant, the time dependence of temperature causes the diffusion coefficient to be time dependent too. Therefore, the natural generalization of the diffusion coefficient for nonequilibrium systems is the diffusivity 
\begin{equation}
\left\langle \left( \Delta r(t) \right)^2  \right\rangle =
6\, \int^t D(t^{\prime}) dt^{\prime}\,.
\label{Dgengen}
\end{equation}
The brackets $\langle \cdots \rangle$ denote averaging over the non-equilibrium ensemble, which evolution is described by a time-dependent $N$-particle distribution function $\rho(t)$ (for simplicity we left the same notation as for the equilibrium average). 

\section{The Pseudo--Liouville Operator}

To describe the dynamics of the granular gas  we use the formalism of the pseudo-Liouville operator ${\cal L}$~\cite{Ernst:69,resibua}. This formalism allows to treat the dynamics of a system of particles with hard-core (i.e. singular) interactions formally in the same way as if the particles would interact via a smooth potential. In classical mechanics the time derivative of any dynamical variable $B$ reads
\begin{equation}
\frac{d}{dt} B\left( \left\{ \vec{r}_i, \vec{v}_i  \right\}, t \right)= \left\{ B , {\cal H} \right\} = i {\cal L} B\,,
\label{Ldef}
\end{equation}
where $\{ \ldots , {\cal H} \}$ denotes the Poisson brackets, which imply differentiation of the Hamiltonian with respect to the coordinates. For singular inter-particle potentials, like a hard-core potential, Poisson brackets are not well defined. Nevertheless a Liouville-like operator may be defined: 
\begin{equation}
i{\cal L}=\sum_j \vec{v}_j \cdot \frac{\partial}{\partial \vec{r}_j}+\sum_{i<j}\, \hat{T}_{\mbox{\footnotesize\em ij}}\,.
\label{L}
\end{equation}
The first sum in (\ref{L}) refers to the free streaming of the particles (the ideal part), while the second sum refers to the interactions of pairs of colliding particles $\{i,j \}$ which are described by the binary collision operator~\cite{Chandler}:
\begin{equation}
\hat{T}_{\mbox{\footnotesize\em ij}}=\sigma^{2} \int d^2\vec{e}\, \Theta \left(- \vec{v}_{\mbox{\footnotesize\em ij}} \cdot \vec{e}\, \right)|\vec{v}_{\mbox{\footnotesize\em ij}} \cdot \vec{e}\, | \delta \left( \vec{r}_{\mbox{\footnotesize\em ij}}- \sigma \vec{e} \right)\left(\hat{b}_{\mbox{\footnotesize\em ij}}^{\vec{e}}-1 \right)  \,.
\label{Tij}
\end{equation}
Here $\sigma=2R$ is the diameter of particles and $\vec{v}_{\mbox{\footnotesize\em  ij}}\cdot\vec{e}=\left(\vec{v}_{i}-\vec{v}_{j}\right)\cdot\vec{e}$ is the normal component of the relative velocity of the colliding pair (which, multiplied by the infinitesimal time $dt$, gives the length of the collision cylinder). The Heaviside function $\Theta(x)$ selects approaching particles and the $\delta$-function represents the hard-core interaction. The operator $\hat{b}_{\mbox{\footnotesize\em ij}}^{{\vec{e}}}$ is defined as
\begin{equation}
\hat{b}_{\mbox{\footnotesize\em ij}}^{\vec{e}} f \left (\vec{r}_{i},  \vec{r}_{j}, \vec{v}_{i},\vec{v}_{j} \cdots \right)=f \left  (\vec{r}_{i}, \vec{r}_{j},  \vec{v}^{\,\prime}_{i},\vec{v}^{\,\prime}_{j} \cdots \right) \, , 
\end{equation}
where $f$ is some function of dynamical variables.  The after-collision velocities of the colliding particles, $\vec{v}^{\,\prime}_{i}$ and $\vec{v}^{\,\prime}_{j}$ are related to their pre-collision values $\vec{v}_{i}$, $\vec{v}_{j}$ via Eq.~(\ref{eq:restdef}).

The pseudo-Liouville operator allows to perform calculations in a very elegant way: Formal integration of Eq. (\ref{Ldef}) yields
 for ($t>t^{\prime}$):
\begin{equation}
B\left( \{ \vec{r}_i, \vec{v}_i  \}, t \right)= e^{i{\cal L} (t-t^{\,\prime} )} B\left( \{     \vec{r}_i, \vec{v}_i \}, t^{\,\prime} \right)\,.
\label{evolA}
\end{equation}
With (\ref{evolA}) the time-correlation function reads
\begin{equation}
\left< B(t^{\prime})B(t) \right >=\int d\Gamma \rho(t^{\prime}) B(t^{\prime}) e^{i{\cal L} (t-t^{\prime})} B(t^{\prime})\,,
\label{evolAA}
\end{equation}
where $\int d\Gamma$ denotes integration over all degrees of freedom and $\rho(t^{\prime})$ depends on temperature $T$, particle number density $n$ , etc., which change on a time-scale $t \gg \tau_c$. In accordance with the molecular chaos assumption at $ t \sim \tau_c$ the sequence of successive collisions occurs without correlations. If $B$ does not depend on the positions of the particles, its time-correlation function reads~\cite{Chandler1}
\begin{equation}
\left\langle B(t^{\prime})B(t) \right \rangle =\left< B^2 \right>_{t^{\prime}} e^{\left.-\left|t-t^\prime\right|\right/\tau_B(t^\prime)}~~~~\left(t > t^{\prime}\right) \,,
\label{AAexp}
\end{equation}
where $\langle \cdots \rangle_{t^\prime}$ denotes averaging with the distribution function taken at time $t^{\prime}$.  The relaxation time $\tau_B$ is inverse to the initial slope of the VAF~\cite{Chandler1}. It may be found from the time derivative of $\left\langle B(t^{\prime} )B(t) \right \rangle$ taken at $t=t^{\prime}$. Eqs.~(\ref{evolAA}) and (\ref{AAexp}) then yield 
\begin{equation}
-\tau_B^{-1}(t^{\prime})=\int d\Gamma \rho(t^{\prime}) B {i{\cal L} }B \left/\left\langle B^2 \right.\right \rangle_{t^{\prime}}= \frac{\left\langle B i{\cal L}  B \right \rangle_{t^{\prime}}}{\left\langle B^2 \right \rangle_{t^{\prime}}}\,.
\label{ALA}
\end{equation}
The relaxation time $\tau_B^{-1}(t^{\prime})$ which depends on time via the distribution function $\rho(t^{\prime})$, changes on the time-scale $t \gg \tau_c$.

\section{Velocity Autocorrelation Function\protect\newline and Self--Diffusion Coefficient}
\subsection{Constant Restitution Coefficient $\epsilon$}

Following the idea described in section \ref{BP:GGD} we calculate the velocity autocorrelation function and the self-diffusion coefficient. In the previous section we discussed a method to calculate the evolution of a general dynamic function $B(t)$. Now we specify $B(t)$ to be the velocity of a tagged particle, say $\vec{v}_1(t)$. Then with $3T(t)=\left\langle v^2 \right \rangle_t$ Eqs.~(\ref{AAexp},\ref{ALA}) (with (\ref{L},\ref{Tij})) read
\begin{equation}
\left\langle \vec{v}_1 (t^{\prime})\cdot  \vec{v}_1(t)\right\rangle = 3T(t^{\prime}) e^{-|t-t^{\prime}|/\tau_v(t^{\prime})}
\label{vvexp}
\end{equation} 
\begin{equation}
-\tau_v^{-1}(t^\prime)=
(N-1) \frac{\left< \vec{v}_1 \cdot \hat{T}_{12} \vec{v}_1\right>_{t^{\prime}}}{\left< \vec{v}_1  \cdot \vec{v}_1 \right>_{t^{\prime}}}\,.
\label{vTv}
\end{equation} 
To obtain (\ref{vTv}) we take into account that ${\cal L}_0 \vec{v}_1=0$, $\hat{T}_{\mbox{\footnotesize\em ij}}\vec{v}_1=0$ (for $i \neq 1$) and the identity of the particles.  The calculation of $\tau_v^{-1}(t^\prime)$ may be performed if we assume that the distribution function $\rho(t^{\prime})$ is a product of the coordinate part, which corresponds to a uniform and isotropic system, and the velocity part being a product of Maxwellian distribution functions 
\begin{equation}
\phi(\vec{v}_i)=\frac{\exp[-v_i^2/2T(t^{\prime})]}{[2\pi T(t^{\prime})]^{3/2}}\,, \quad i=1, \ldots, N\, .  
\end{equation}
We want to mention that a more sophisticated analysis (e.g. \cite{BrilliantovPoeschelPRE200:vdf,BrilliantovPoeschelPRE200:vdfeps(v)}) shows that the velocity distribution function deviates from the Maxwellian. In this paper we neglect these deviations which are small for small inelasticity of the particles ($\epsilon\rightarrow 1$). Calculation of the diffusion coefficient with non-Maxwellian velocity distribution function has been performed in \cite{BrilliantovPoeschelProceed:2000}. The result (depending on the inelasticity) may differ quantitatively from that given here, but its  functional form  remains unaffected.  
 
Integration over the coordinate part in (\ref{vTv}) yields
\begin{equation}
(N-1)\int\rho(t^{\prime}) \delta \left(\vec{r}_{\mbox{\footnotesize\em ij}}-\sigma \vec{e} \right) d\vec{r}_1 \cdots d\vec{r}_N\
=ng_2(\sigma) \prod_i \phi(\vec{v}_i )\,,  
\end{equation}
where we use the definition of the two-particle distribution function and where
\begin{equation}
g_2(\sigma)=\frac12 (2-\eta)/(1-\eta)^3  
\end{equation}
is its contact value~\cite{resibua,Chandler1}, which depends on the packing fraction
$\eta=\frac16 \pi n\sigma^3 $. With 
\begin{equation}
\left< \vec{v}_1 \hat{T}_{12} \vec{v}_1 \right>_{t^{\prime}} =\frac12 \left< \vec{v}_{12} \hat{T}_{12} \vec{v}_1 \right>_{t^{\prime}}   
\end{equation}
due to the collision rules and definition (\ref{Tij}) one finally arrives at 
\begin{equation}
\label{vTv2}
\tau_v^{-1}(t^\prime) = \!\frac{ng_2(\sigma) \sigma^{2}}{4}\!\! \int\!\! d\vec{v}_{12} \phi (\vec{v}_{12} ) \!\!\int\!\! d^2 \vec{e} \frac{  \Theta(-\vec{v}_{12} \cdot \vec{e}) |\vec{v}_{12} \cdot \vec{e}|\,(\vec{v}_{12} \cdot \vec{e})^2(1+\epsilon)}{\left< \vec{v}_1 \cdot  \vec{v}_1\right >_{t^{\prime}}}
\end{equation}
where 
\begin{equation}
\phi (\vec{v}_{12} )=(4\pi T)^{-3/2} \exp(-v_{12}^2/4T)  
\end{equation}
is the Maxwellian distribution for the relative velocity of two particles. For $\epsilon$ not depending on $v_{12}$ Eq.~(\ref{vTv2}) yields
\begin{equation}
\tau_v^{-1}(t)=\frac{\epsilon +1}{2}\frac83 n \sigma^2 g_2(\sigma) \sqrt{\pi T(t)} =\frac{\epsilon +1}{2}  \tau_E^{-1}(t)\,, 
\label{tEt}
\end{equation}
where $\tau_E(t)=\frac32\,\tau_c(t)$ is the Enskog relaxation time~\cite{resibua}. For the granular gas it depends on time according to the same time-scale as temperature. According to Eq.~(\ref{tEt}) the velocity correlation time for inelastic collisions is larger than for elastic collisions. This follows from partial suppression of backscattering of particles due to inelastic losses in their normal relative motion. As a result for inelastic particle interaction the angle between crossing trajectories after a collision is smaller than before the collision, while for elastic interaction both angles are identical. Therefore, the velocity correlation time is larger for inelastically colliding particles.

Using the velocity correlation function one writes
\begin{equation}
\left\langle \left( \Delta r(t) \right)^2  \right\rangle =2 \int_0^t dt^\prime 3T (t^\prime) \int_{t^\prime}^t dt^{\prime\prime} e^{-|t^{\prime\prime}-t^\prime|/\tau_v(t^\prime)}\,.
\label{Difvel}
\end{equation}
On the short-time scale $t \sim \tau_c$, $T (t^\prime)$ and $\tau_v(t^\prime)$ may be considered as constants. Integrating in (\ref{Difvel}) over $t^{\prime\prime}$ and equating with (\ref{Dgengen}) yields for $t \gg \tau_c \sim \tau_v$ the diffusivity (time-dependent self-diffusion coefficient)
\begin{equation}
D(t)= T(t) \tau_v(t)\,.
\label{Difviatau}
\end{equation}

Using the pseudo-Liouville operator one can also describe the time-depen\-dence of the granular temperature: From (\ref{Ldef}) it follows (see also \cite{HuthmannZippelius:97}):
\begin{equation}
\dot{T}(t)=\frac13\, \frac{d}{dt} \left< v^2 \right>_t=\frac13\,  \left< i {\cal L} \,v^2 \right>_t \, .  
\end{equation}
Calculations similar to that for  $\tau_v(t)$ yield
\begin{equation}
\label{Tepsconstder}
\dot{T}(t)= -\left[ \frac43 \pi^{1/2}g_2(\sigma)\sigma^2 n (1-\epsilon^2) \right]\, T^{3/2} \, .  
\end{equation}
So that finally one obtains a well known result for temperature
\begin{equation}
\label{Tepsconst}
T(t)/T_0=\left[ 1+\gamma_0 t/\tau_c(0) \right]^{-2}\,,  
\end{equation}
where $\gamma_0 \equiv (1-\epsilon^2)/6$ \cite{GZ93,NEBO97}, $T_0$ is the initial temperature at $t=0$, and 
\begin{equation}
\tau_c(0) = 4\pi^{1/2}g_2(\sigma)\sigma^2 n \sqrt{T_0} 
\end{equation}
is the  initial mean collision time.  
Thus, using Eqs.~(\ref{tEt}) and  (\ref{Difviatau}) one obtains for the self-diffusion coefficient for the case $\epsilon=\mbox{const.}$
\begin{equation}
D(t)= \left( \frac{2}{\epsilon +1} \right) D_0 
\left[ 1+\gamma_0 t/\tau_c(0) \right]^{-1}
\label{Difviatau1}
\end{equation}
where $D_0$ is the Enskog self-diffusion coefficient for elastic particles:

\begin{equation}
\label{DEnskog}
D_0^{-1}= \frac83 \pi^{1/2}\sigma^2 g_2(\sigma) n  T_0^{-1/2} \, .
\end{equation}
Correspondingly, the mean-square displacement reads for $t \gg \tau_c(0)$:\\[-0.1cm]
\begin{equation}
\label{drcons}
\left< \left( \Delta r(t) \right)^2 \right>  \sim \log t \, .
\end{equation}
\vspace*{-0.8cm}

\subsection{Impact Velocity Dependent Restitution Coefficient}

As mentioned before the assumption of a constant restitution coefficient contradicts basic physical understanding and does not even agree with a dimension analysis~\cite{Rozaetal}. Nevertheless it was used in many publications for simplicity of the calculation. In this section we will calculate the diffusivity for a gas of viscoelastic particles for which $\epsilon$ is given by Eq. (\ref{epsilon}) and, surprisingly, we will find a qualitatively different result.

The velocity correlation time $\tau_v$ can be found from Eq.~(\ref{vTv2}) where $\epsilon$ depends on the impact velocity as described by Eq.~(\ref{epsilon}):
\begin{eqnarray}
\tau_v^{-1}(t)&=&\tau_E^{-1}(t) \left[1-\frac34 \Gamma\left( \frac{21}{10}  \right) C_1 A \alpha^{2/5}\left(4T(t) \right)^{1/10} +\right.\nonumber \\
&+& \left. \frac{27}{40}\Gamma\left( \frac{11}{5} \right) C_1^2 A^2 \alpha^{4/5} \left( 4T(t) \right)^{1/5} \mp \cdots \right]\,,
\label{tfin}
\end{eqnarray}
where $\Gamma(x)$ is the Gamma-function, $\tau_E$ is given by Eq.~(\ref{tEt}) and we use Eq.~(\ref{C2}), which relates the coefficients $C_1$ and $C_2$.  

Performing calculations similar to those for the constant restitution coefficient, one obtains for the time evolution of temperature 
\begin{equation}
\dot{T}/T_0=-(5/3)\tau_0^{-1} \left[ (T/T_0)^{8/5}-b\, \delta \, (T/T_0)^{17/10} + \ldots \right] 
\label{Tdot1}
\end{equation}
where we introduce the small (dissipative) parameter
\begin{equation}
\delta=A\alpha^{2/5}T_0^{1/10} \, ,   
\end{equation}
the characteristic time for the temperature evolution
\begin{equation}
\tau_0^{-1} = \delta \cdot \tau_c(0)^{-1} \cdot \frac35 \, 2^{1/5} \,C_1 \Gamma\left( \frac{21}{10} \right) \approx 0.831928 \cdot \delta \cdot \tau_c(0)^{-1}  
\end{equation}
and the numerical constant $b \equiv (18/5)2^{1/5}C_1\Gamma(1/5)/\Gamma(1/10)\approx 2.3018$. Solving Eq.~(\ref{Tdot1}) and expanding the result in terms of the small parameter $\delta$ yields
\begin{equation}
\frac{T(t)}{T_0}=\left(1+\frac{t}{\tau_0} \right)^{-\frac{5}{3}} + a_1 \, \delta \left(1+\frac{t}{\tau_0}\right)^{-\frac{11}{6}}+ a_2 \delta^2 \, \left(1+\frac{t}{\tau_0}\right)^{-2} + \cdots
\label{Tres}
\end{equation}
with $a_1$ and $a_2$ being pure numbers~\cite{constants}.

 From Eqs.~(\ref{Difviatau},\ref{Tres},\ref{tfin}) follows for the self-diffusion coefficient:
\begin{equation}
\frac{D(t)}{D_0}= \left(1+\frac{t}{\tau_0}
\right)^{-\frac{5}{6}} + a_3\,\delta 
\left(1+\frac{t}{\tau_0}\right)^{-1}+ a_4 
\delta^2 \,\left(1+\frac{t}{\tau_0}
\right)^{-\frac{7}{6}} + \cdots\,,
\label{Dres}
\end{equation}
where $D_0$ is given in Eq.~(\ref{DEnskog}) and $a_3$, $a_4$ are pure numbers~\cite{constants}. Correspondingly, the mean-square displacement reads for $\tau_0 \ll t$:
\begin{equation}
\left\langle \left( \Delta r(t) \right)^2  \right\rangle  \sim 
t^{1/6}+a_3 \delta \,\log t \, .
\label{dRasym}
\end{equation}
This dependence holds true for 
\begin{equation}
\tau_c(0)\,\delta^{-1} \ll t \ll \tau_c(0)\, \delta^{-11/5}\,,   
\label{Gl44}
\end{equation}
where the
first inequality follows from the condition $\tau_0 \ll t$, while the
second one follows from the condition $\tau_c(t) \ll
\tau_0$ which makes the concept of the temperature meaningful.  

Comparing Eqs.~(\ref{dRasym}) and (\ref{drcons}) one notes that the impact-velocity dependent restitution coefficient (\ref{epsilon}) leads to a significant change of the long-time behavior of the mean-square displacement of particles in time. Compared to its logarithmically weak dependence for $\epsilon=\mbox{const.}$ (simplified collision model), the impact-velocity dependence of the restitution coefficient (\ref{epsilon}) as it appears for viscoelastic particles gives rise to a considerably faster spreading of particles according to a power law.

\vspace*{-0.2cm}
\section{Results and Discussion}
\vspace*{-0.2cm}

We studied the diffusion of particles in a homogeneously cooling granular gas. With the assumption of molecular chaos we calculated the velocity time-correlation function and the time dependent self-diffusion coefficient (diffusivity) $D(t)$. For the case of constant coefficient of restitution $\epsilon=\mbox{const.}$ the diffusivity $D(t)$ is expressed in terms of the model parameter $\epsilon$. For the more realistic case of an impact velocity dependent restitution coefficient, as it has been derived for viscoelastic particle interaction, we found a relation which expresses the diffusivity in terms of material constants of the particles and characteristics of the granular gas, such as temperature, density, etc. 

For granular particles suffering viscoelastic collisions we found that the mean-square displacement grows with time as a power law $\left\langle \left( \Delta r(t) \right)^2  \right\rangle  \sim t^{1/6}$, i.e. much faster than the logarithmic growth $\left\langle \left( \Delta r(t) \right)^2  \right\rangle \sim \log t$ for the case of a constant restitution coefficient. It worth to note that qualitatively this power-law dependence (as well as the logarithmic one) simply follows from scaling arguments and the time-dependence of temperature. Indeed, the average velocity scales as $\bar{v} \sim T^{1/2}$, and therefore as $\sim t^{-1}$ for the constant restitution coefficient and as $\sim t^{-5/6}$ for the impact-velocity dependent coefficient. The diffusivity in granular gas scales as $D \sim l^2/\tau_c$, where $l \sim \sigma^{-2} n^{-1}$ is the mean-free path, which does not change with time (in the regime preceding clustering), and $\tau_c \sim l/\bar{v}$ is the mean-collision time. Thus, $D \sim l \bar{v} \sim T^{1/2}$, and we obtain that the mean-square displacement, $\int^t D(t)dt$, scales as $\sim \log t$ in the former case and as $\sim t^{1/6}$ in the latter. 

What will be the impact of this apparently dramatic difference in the time dependence of $\left< \left( \Delta r (t) \right)^2 \right> $ for the properties of granular gases? Obviously, viscoelastic particles spread (and, therefore, mix) quicker than particles which interact via $\epsilon=\mbox{const.}$ Since the temperature decreases more slowly for the former case, as $\sim t^{-5/3}$, as compared with $\sim t^{-2}$ for the latter one, retarded clustering may be expected. Let us explain this in more detail: According to the linear stability analysis~\cite{GZ93,clus2}, long-range density fluctuations in homogeneous granular gas occur to be unstable, i.e. they grow exponentially with time, leading to cluster formation. The critical value of the stable wavelengths $\lambda_c = 2\pi/k_c$ ($k$ is the wave number) reads $\lambda_c \sim l/\sqrt{1-\epsilon^2}$~\cite{GZ93}, see also~\cite{clus2}. All density fluctuations with $\lambda > \lambda_c$ are unstable. For the case of viscoelastic particles the effective value of $\epsilon$ permanently grows with time (as temperature decreases), and $\lambda_c$ grows accordingly. This implies that the size of the regions, where the gas is still homogeneous, grows with time, since density fluctuations do not blow up on length scales smaller 
than $\lambda_c$. Thus for the velocity-dependent restitution coefficient the critical 
wavelength increases with time (i.e. one has a time-dependent criterion for the size of the region which is unstable in the HCS). As compared to the case $\epsilon={\rm const.}$, clustering occurs on a larger length-scale; hence the conditions to remain in the homogeneous medium persist for a longer time for viscoelastic particles. 

Obviously, self-diffusion directly counteracts clustering. Therefore, we conclude, that for a granular gas of viscoelastic particles which implies an impact velocity dependent restitution coefficient, clustering is retarded as compared with a gas under the oversimplified assumption $\epsilon=\mbox{const.}$ 

\medskip

\noindent We thank M. H. Ernst and I. Goldhirsch for valuable discussions.


\begin{thebibliography}{7}
\bibitem{Chandler} For application to ``ordinary'' fluids, see~\cite{Chandler1} and to granular systems~\cite{NoijeErnst:97,HuthmannZippelius:97}. A rigorous definition of ${\cal L}$ includes a pre-factor, preventing successive collisions of pairs of particles~\cite{Ernst:69,resibua} which, however, does not affect the present analysis.
\end{thebibliography}
\end{document}